\documentclass[12pt]{article}

\usepackage[totalheight = 23cm, totalwidth = 17cm]{geometry}
\usepackage{amssymb,amsmath,amsfonts,amsbsy,graphicx,simplewick}

\newcommand{\mpl}{m_{\rm Pl}}
\newcommand{\calP}{{\cal P}}
\newcommand{\calR}{{\cal R}}
\newcommand{\fnl}{f_{\rm NL}}
\newcommand{\calL}{{\cal L}}

\begin{document}

\begin{titlepage}

\begin{center}

\vspace*{-10ex}
 \hspace*{\fill} ICTS-USTC-09-21

\vskip 1.cm

\Huge{Non-Gaussianity from false vacuum inflation: Old curvaton scenario}

\vskip 1cm

\large{
 Jinn-Ouk Gong$^{\dag}$\footnote{jgong@lorentz.leidenuniv.nl}
 \hspace{0.2cm}
 Chunshan Lin$^{\S,\P}$\footnote{lics@mail.ustc.edu.cn}
 \hspace{0.2cm}
 Yi Wang$^{\P}$\footnote{wangyi@hep.physics.mcgill.ca}
\\
\vspace{0.5cm}
 {\em
 ${}^\dag$Instituut-Lorentz for Theoretical Physics, Universiteit Leiden
 \\
 Niels Bohrweg 2, 2333 CA Leiden, The Netherlands
 \\
 \vspace{0.2cm}
 ${}^\S$Interdisciplinary Center of Theoretical Studies
 \\
 University of Science and Technology of China
 \\
 Hefei, Anhui 230026, China
 \\
 \vspace{0.2cm}
 ${}^\P$Department of Physics, McGill University
 \\
 Montr\'{e}al, QC, H3A 2T8, Canada}
}

\vskip 0.5cm

\today

\vskip 1.2cm

\end{center}

\begin{abstract}

We calculate the three-point correlation function of the comoving curvature perturbation
generated during an inflationary epoch driven by false vacuum energy. We get a novel
false vacuum shape bispectrum, which peaks in the equilateral limit. Using this result,
we propose a scenario which we call ``old curvaton''. The shape of the resulting
bispectrum lies between the local and the false vacuum shapes. In addition we have a
large running of the spectral index.

\end{abstract}

\end{titlepage}

\setcounter{page}{0}
\newpage
\setcounter{page}{1}

\section{Introduction}

Inflation~\cite{Guth:1980zm,newinflation} nowadays is accepted as the standard paradigm
to solve many cosmological problems, and to provide the appropriate initial conditions
for the hot big bang cosmology. One of the strongest supports of this early era of
inflation is the observed primordial curvature perturbation $\calR_c$, whose power
spectrum $\calP_\calR$ is nearly scale invariant with its index being $n_\calR \approx
0.96$, with almost perfect Gaussian statistics~\cite{Komatsu:2008hk}. Conversely, these
observations have been acting as a powerful discriminator among the models of inflation
based on particle physics~\cite{Lyth:1998xn} and placing strong constraints on the
realization of inflation, e.g. in the context of supergravity~\cite{Copeland:1994vg}. In
the next few years, we will obtain more precise data on the primordial density
perturbations by ongoing and forthcoming experiments like the Wilkinson Microwave
Anisotropy Probe (WMAP) and the Planck satellites, and thus will be able to constrain the
models of inflation even stronger than now.

In the light of these upcoming observations, there have been increasing interests in the
non-Gaussian signature of the primordial perturbations. Currently, the non-Gaussianity of
the cosmic microwave background (CMB) fluctuations is constrained to be $|\fnl| \lesssim
100$~\cite{Komatsu:2008hk}, with $\fnl$ being the non-linear
parameter~\cite{Komatsu:2001rj}, which translates into less than 0.1\% of non-Gaussian
contribution. The sensitivity is expected to be improved at the level of $\fnl \sim
\mathcal{O}(1)$ on the CMB scales \cite{bluebook}, and indeed primordial non-Gaussianity
will be another powerful probe of inflation: if substantial non-Gaussianity were to be
detected, we should look for new models beyond the simplest single field slow-roll
inflation models, where $\fnl \ll{\cal O} (1)$~\cite{Maldacena:2002vr}. A short list of
the models where large non-Gaussianity may arise includes Dirac-Born-Infeld (DBI)
inflation~\cite{dbi}, single field inflation with features\footnote{In this case, we also
have characteristic modulation and oscillation in the power
spectrum~\cite{featurespectrum}.}~\cite{featurenG}, multi-field inflation\footnote{Note
also an intermediate model of quasi-single field
inflation~\cite{quasiSingleFieldNG}.}~\cite{multifieldnG}, curvaton~\cite{curvatonnG} ,
and so on.

In this note, we study a new possibility of generating large amount of highly
non-Gaussian curvature perturbation: we consider a period of inflation supported by a
non-zero vacuum energy. This is in fact the original model of
inflation~\cite{Guth:1980zm} and has received a renewed interest in the context of vast
string landscape~\cite{newoldinflation}. But $\calR_c$ generated during this stage by
vacuum fluctuations has been unknown until very recently~\cite{Gong:2008ni}. The
difficulty is that while the inflaton field $\phi$ is well anchored in a false vacuum,
the comoving curvature perturbation, which is given by
\begin{equation}\label{rcprev}
\calR_c \sim \frac{H}{\dot\phi}\delta\phi \, ,
\end{equation}
is ill defined. The origin of this difficulty is that during false
vacuum inflation the inflaton $\phi$ is classically not moving,
and there is no preferred time direction to define comoving
hypersurfaces. But this phase should be broken, and indeed in
Ref.~\cite{Gong:2008ni} $\calP_\calR$ was calculated by
incorporating a regularizing mass which scales as $a^{-2}$. The
results are such that $\calP_\calR$ is dependent on the scale of
the regularizing mass, with steeply blue index $n_\calR =
4$~\cite{Gong:2008ni} and nearly $\chi^2$
distribution\footnote{Note that in Ref.~\cite{Yamamoto:1995sw} a
similar situation was discussed, with the power spectrum induced
from the false vacuum inflation and the quantum tunneling in
spatially open universe. We expect similar results, i.e.
$\calP_\calR$ is suppressed on large scales and is highly scale
dependent with $n_\calR = 4$, since these are related to the
properties of heavy fields in de Sitter space."}~\cite{prep}.
Thus, we need an additional mechanism to generate $\calR_c$ to
match observations.

In addition to false vacuum inflation, we incorporate the curvaton
mechanism~\cite{curvaton} to generate the Gaussian contribution of $\calR_c$. In this
combined scenario of false vacuum inflation and the curvaton mechanism, or shortly ``old
curvaton'' scenario, as we will see we can build a concrete model with observationally
consistent $\calP_\calR$, while providing substantial degree of non-Gaussian signature.
Especially, the bispectrum lies between the so-called equilateral and local types, giving
a concrete example of mixed non-Gaussianity. The structure of this note is as follows. In
Section~\ref{sec_3pointcorr}, we explicitly calculate the three-point correlation
function, or the bispectrum of $\calR_c$ using the results of Ref.~\cite{Gong:2008ni}.
Then in Section~\ref{sec_oldcurvaton} we incorporate the curvaton mechanism and present
the resulting power spectrum and the bispectrum in the old curvaton scenario. We conclude
in Section~\ref{sec_conclusion}. Some technical detail is given in Appendix.

\section{Bispectrum from false vacuum inflation}
\label{sec_3pointcorr}

We consider a theory where the inflaton sector is described by the Einstein gravity with
scalar Lagrangian,
\begin{equation}
\calL = \frac{\mpl^2}{2}R - \frac{1}{2}\phi^{,\mu}\phi_{,\mu} - V(\phi) \, ,
\end{equation}
where $\mpl = (8\pi G)^{-1/2} \approx 2.4 \times 10^{18}\mathrm{GeV}$. The potential is
assumed to have the form
\begin{equation}\label{potential}
V(\phi) = V_0 + \frac{1}{2} \left( -m_\phi^2 + \frac{\mu^2}{a^2} \right)\phi^2 \, .
\end{equation}
We are interested in the epoch while $m_\mathrm{eff}^2 = -m_\phi^2 + \mu^2/a^2 > 0$ so
that $\phi$ is well anchored at $\phi = 0$. During this stage, as shown in Appendix
\ref{appendixA}, we can find the comoving curvature perturbation ${\cal R}_c$ as
\begin{equation}\label{falsevacuum_Rc}
\calR_c(\mathbf{k}) =
\frac{1}{2a^2\langle\rho+p\rangle_\mathrm{ren}(\mu\eta)^2} \int
\frac{d^3q}{(2\pi)^3}\phi_\mathbf{q}\phi_{\mathbf{k}-\mathbf{q}}
\left[ \mu^2 + \mathbf{q}\cdot(\mathbf{q}-\mathbf{k}) \right] \, .
\end{equation}
The convolution in the momentum space will lead to a loop integral
if we calculate the two-point or three-point correlation function.
It's worth to note that this kind of loop integral is not the real
loop integral in interactional quantum field theory. It is the
so-called c-loop~\cite{Lyth:2006qz}.

As can be read from (\ref{falsevacuum_Rc}), the curvature
perturbation is quadratic in $\delta\phi$, so we need not follow
the conventional approach [8]. The bispectrum arise from c-loop
integration between different modes of curvature perturbation. The
three-point correlation function of $\mathcal{R}_c$ is explicitly
written as
\begin{align}\label{falsevacuum_3point}
\left\langle
\mathcal{R}_c(\mathbf{k}_1)\mathcal{R}_c(\mathbf{k}_2)\mathcal{R}_c(\mathbf{k}_3)
\right\rangle = & \frac{1}{\left[
2a^2\langle\rho+p\rangle_\mathrm{ren}(\mu\eta)^2 \right]^3} \int
\frac{d^3q_1}{(2\pi)^3}\frac{d^3q_2}{(2\pi)^3}\frac{d^3q_3}{(2\pi)^3}
\left\langle \phi_{\mathbf{q}_1}\phi_{\mathbf{k}_1-\mathbf{q}_1}
\phi_{\mathbf{q}_2}\phi_{\mathbf{k}_2-\mathbf{q}_2}
\phi_{\mathbf{q}_3}\phi_{\mathbf{k}_3-\mathbf{q}_3} \right\rangle
\nonumber\\
& \times \left[ \mu^2 +
\mathbf{q}_1\cdot(\mathbf{q}_1-\mathbf{k}_1) \right] \left[ \mu^2
+ \mathbf{q}_2\cdot(\mathbf{q}_2-\mathbf{k}_2) \right] \left[
\mu^2 + \mathbf{q}_3\cdot(\mathbf{q}_3-\mathbf{k}_3) \right] \, .
\end{align}
We can after some straightforward calculations find the expectation value by explicitly
plugging (\ref{phi_decomposition}) into (\ref{falsevacuum_3point}), but we can anticipate
the result by considering the possible contractions between $\phi$'s. It is obvious that
there are 8 possible physically meaningful contractions: $\calR_c(\mathbf{k}_1)$ to
$\calR_c(\mathbf{k}_2)$ or $\calR_c(\mathbf{k}_3)$, so we have 2 choices. Further, since
each $\calR_c(\mathbf{k}_i)$ has 2 field contents, there are 2 choices for each: one of
the fields in (say) $\calR_c(\mathbf{k}_1)$ can be correlated to one in (say)
$\calR_c(\mathbf{k}_2)$ and the other in $\calR_c(\mathbf{k}_1)$ can be correlated to one
in $\calR_c(\mathbf{k}_3)$. This is to avoid self-contraction of $\calR_c(\mathbf{k}_3)$.
Thus total $2\times2\times2=8$ different ways of contraction. These are explicitly
written as
\begin{align}
&
\contraction{\langle(}{\phi}{{}_{\mathbf{k}_1}\phi_{\mathbf{k}_1-\mathbf{q}_1})(}{\phi}
\contraction[2ex]{\phi_{\mathbf{k}_1}}{\phi_{\mathbf{k}_1-}}{{}_{\mathbf{q}_1})(\phi_{\mathbf{k}_2}\phi_{\mathbf{k}_2-\mathbf{q}_2})(\phi}{{}_{\mathbf{k}_3}}
\contraction{(\phi_{\mathbf{k}_1}\phi_{\mathbf{k}_1-\mathbf{q}_1})(\phi_{\mathbf{k}_2}}{\phi_{\mathbf{k}}}{{}_{{}_2-\mathbf{q}_2})(\phi_{\mathbf{k}_3}}{\phi_{\mathbf{k}_3}}
\langle
(\phi_{\mathbf{k}_1}\phi_{\mathbf{k}_1-\mathbf{q}_1})(\phi_{\mathbf{k}_2}\phi_{\mathbf{k}_2-\mathbf{q}_2})(\phi_{\mathbf{k}_3}\phi_{\mathbf{k}_3-\mathbf{q}_3})
\rangle +
\contraction{\langle(}{\phi}{{}_{k_1}\phi_{|\mathbf{k}_1-\mathbf{q}_1|})(}{\phi}
\contraction[2ex]{\phi_{k_1}}{\phi_{|\mathbf{k}_1-}}{{}_{\mathbf{q}_1|})(\phi_{k_2}\phi_{|\mathbf{k}_2-\mathbf{q}_2|})(\phi_{k_3}}{\phi_{\mathbf{k}_3-}}
\contraction{(\phi_{k_1}\phi_{|\mathbf{k}_1-\mathbf{q}_1|})(\phi_{k_2}}{\phi_{|\mathbf{k}}}{{}_{{}_2-\mathbf{q}_2|})(}{\phi_{k_3}}
\langle
(\phi_{\mathbf{k}_1}\phi_{\mathbf{k}_1-\mathbf{q}_1})(\phi_{\mathbf{k}_2}\phi_{\mathbf{k}_2-\mathbf{q}_2})(\phi_{\mathbf{k}_3}\phi_{\mathbf{k}_3-\mathbf{q}_3})
\rangle
\nonumber\\
+ &
\contraction{(}{\phi_{k_1}}{\phi_{|\mathbf{k}_1-\mathbf{q}_1|})(\phi_{k_2}}{\phi_{\mathbf{k}_2}}
\contraction[2ex]{\phi_{k_1}}{\phi_{|\mathbf{k}_1-}}{{}_{\mathbf{q}_1|})(\phi_{k_2}\phi_{|\mathbf{k}_2-\mathbf{q}_2|})(\phi}{{}_{k_3}}
\contraction[3ex]{(\phi_{k_1}\phi_{|\mathbf{k}_1-\mathbf{q}_1|})(}{\phi_{k_2}}{\phi_{|\mathbf{k}_2-\mathbf{q}_2|})(\phi_{k_3}}{\phi_{|\mathbf{k}}}
\langle
(\phi_{\mathbf{k}_1}\phi_{\mathbf{k}_1-\mathbf{q}_1})(\phi_{\mathbf{k}_2}\phi_{\mathbf{k}_2-\mathbf{q}_2})(\phi_{\mathbf{k}_3}\phi_{\mathbf{k}_3-\mathbf{q}_3})
\rangle +
\contraction{(}{\phi_{k_1}}{\phi_{|\mathbf{k}_1-\mathbf{q}_1|})(\phi_{k_2}}{\phi_{\mathbf{k}_2}}
\contraction[2ex]{\phi_{k_1}}{\phi_{|\mathbf{k}_1-}}{{}_{\mathbf{q}_1|})(\phi_{k_2}\phi_{|\mathbf{k}_2-\mathbf{q}_2|})(\phi_{k_3}}{\phi_{\mathbf{k}_3-}}
\contraction[3ex]{(\phi_{k_1}\phi_{|\mathbf{k}_1-\mathbf{q}_1|})(}{\phi_{k_2}}{\phi_{|\mathbf{k}_2-\mathbf{q}_2|})(}{\phi_{k_3}}
\langle
(\phi_{\mathbf{k}_1}\phi_{\mathbf{k}_1-\mathbf{q}_1})(\phi_{\mathbf{k}_2}\phi_{\mathbf{k}_2-\mathbf{q}_2})(\phi_{\mathbf{k}_3}\phi_{\mathbf{k}_3-\mathbf{q}_3})
\rangle
\nonumber\\
+ &
\contraction{(}{\phi_{\mathbf{k}_1}}{\phi_{\mathbf{k}_1-\mathbf{q}_1})(\phi_{\mathbf{k}_2}\phi_{\mathbf{k}_2-\mathbf{q}_2})(}{\phi_{\mathbf{k}_3}}
\contraction[2ex]{\phi_{k_1}}{\phi_{\mathbf{k}_1-}}{{}_{\mathbf{q}_1|})(\phi_k}{{}_{{}_2}}
\contraction[2ex]{(\phi_{k_1}\phi_{|\mathbf{k}_1-\mathbf{q}_1|})(\phi_{k_2}}{\phi_{|\mathbf{k}}}{{}_{{}_2-\mathbf{q}_2|})(\phi_{k_3}}{\phi_{\mathbf{k}}}
\langle
(\phi_{\mathbf{k}_1}\phi_{\mathbf{k}_1-\mathbf{q}_1})(\phi_{\mathbf{k}_2}\phi_{\mathbf{k}_2-\mathbf{q}_2})(\phi_{\mathbf{k}_3}\phi_{\mathbf{k}_3-\mathbf{q}_3})
\rangle +
\contraction{(}{\phi_{k_1}}{\phi_{|\mathbf{k}_1-\mathbf{q}_1|})(\phi_{k_2}\phi_{|\mathbf{k}_2-\mathbf{q}_2|})(}{\phi_{k_3}}
\contraction[2ex]{\phi_{k_1}}{\phi_{\mathbf{k}_1-}}{{}_{\mathbf{q}_1|})(\phi_{k_2}\phi}{{}_{\mathbf{k}_2-}}
\contraction[3ex]{(\phi_{k_1}\phi_{|\mathbf{k}_1-\mathbf{q}_1|})(}{\phi_{k_2}}{\phi_{|\mathbf{k}_2-\mathbf{q}_2|})(\phi_{k_3}}{\phi_{\mathbf{k}}}
\langle
(\phi_{\mathbf{k}_1}\phi_{\mathbf{k}_1-\mathbf{q}_1})(\phi_{\mathbf{k}_2}\phi_{\mathbf{k}_2-\mathbf{q}_2})(\phi_{\mathbf{k}_3}\phi_{\mathbf{k}_3-\mathbf{q}_3})
\rangle
\nonumber\\
+ &
\contraction{(}{\phi_{k_1}}{\phi_{|\mathbf{k}_1-\mathbf{q}_1|})(\phi_{k_2}\phi_{|\mathbf{k}_2-\mathbf{q}_2|})(\phi_{k_3}}{\phi_{|\mathbf{k}}}
\contraction[2ex]{\phi_{k_1}}{\phi_{\mathbf{k}_1-}}{{}_{\mathbf{q}_1|})(\phi_k}{{}_{{}_2}}
\contraction[2ex]{(\phi_{k_1}\phi_{|\mathbf{k}_1-\mathbf{q}_1|})(\phi_{k_2}}{\phi_{|\mathbf{k}}}{{}_{{}_2-\mathbf{q}_2|})(}{\phi_{k_3}}
\langle
(\phi_{\mathbf{k}_1}\phi_{\mathbf{k}_1-\mathbf{q}_1})(\phi_{\mathbf{k}_2}\phi_{\mathbf{k}_2-\mathbf{q}_2})(\phi_{\mathbf{k}_3}\phi_{\mathbf{k}_3-\mathbf{q}_3})
\rangle +
\contraction{(}{\phi_{k_1}}{\phi_{|\mathbf{k}_1-\mathbf{q}_1|})(\phi_{k_2}\phi_{|\mathbf{k}_2-\mathbf{q}_2|})(\phi_{k_3}}{\phi_{|\mathbf{k}}}
\contraction[2ex]{\phi_{k_1}}{\phi_{\mathbf{k}_1-}}{{}_{\mathbf{q}_1|})(\phi_{k_2}\phi}{{}_{\mathbf{k}_2-}}
\contraction[3ex]{(\phi_{k_1}\phi_{|\mathbf{k}_1-\mathbf{q}_1|})(}{\phi_{k_2}}{\phi_{|\mathbf{k}_2-\mathbf{q}_2|})(}{\phi_{k_3}}
\langle
(\phi_{\mathbf{k}_1}\phi_{\mathbf{k}_1-\mathbf{q}_1})(\phi_{\mathbf{k}_2}\phi_{\mathbf{k}_2-\mathbf{q}_2})(\phi_{\mathbf{k}_3}\phi_{\mathbf{k}_3-\mathbf{q}_3})
\rangle \, ,
\end{align}
and these terms exactly correspond to the terms with
$\delta^{(3)}(\mathbf{k}_1+\mathbf{k}_2+\mathbf{k}_3)$ which we can obtain by direct
computations. Indeed, after some trivial calculations, we find that
(\ref{falsevacuum_3point}) can be written as
\begin{align}\label{falsevacuum_3point2}
& \left\langle
\mathcal{R}_c(\mathbf{k}_1)\mathcal{R}_c(\mathbf{k}_2)\mathcal{R}_c(\mathbf{k}_3)
\right\rangle
\nonumber\\
& = \frac{4\delta^{(3)}(\mathbf{k}_1+\mathbf{k}_2+\mathbf{k}_3)}{\left[
2a^2\langle\rho+p\rangle_\mathrm{ren}(\mu\eta)^2 \right]^3}
\nonumber\\
& \hspace{0.5cm} \times \int d^3q \left\{
|\varphi_q|^2|\varphi_{|\mathbf{k}_1-\mathbf{q}|}|^2|\varphi_{|\mathbf{k}_2+\mathbf{q}|}|^2
\left[ \mu^2 + \mathbf{q} \cdot \left( \mathbf{q} - \mathbf{k}_1 \right) \right] \left[
\mu^2 + \mathbf{q} \cdot \left( \mathbf{k}_2 + \mathbf{q} \right) \right] \left[
\mu^2+(\mathbf{q}-\mathbf{k}_1) \cdot (\mathbf{k}_2+\mathbf{q}) \right] \right.
\nonumber\\
& \left. \hspace{2.0cm} +
|\varphi_q|^2|\varphi_{|\mathbf{k}_1-\mathbf{q}|}|^2|\varphi_{|\mathbf{k}_3+\mathbf{q}|}|^2
\left[ \mu^2 + \mathbf{q} \cdot \left( \mathbf{q} - \mathbf{k}_1 \right) \right] \left[
\mu^2 + \mathbf{q} \cdot \left( \mathbf{k}_3 + \mathbf{q} \right) \right] \left[
\mu^2+(\mathbf{q}-\mathbf{k}_1) \cdot (\mathbf{k}_3+\mathbf{q}) \right] \right\} \, .
\end{align}
Where $\varphi_k$ is the mode function of perturbation
$\phi_{\mathbf{k}}$ (see (A.11) for example),
\begin{equation}
\phi_{\mathbf{k}}=a_{\mathbf{k}}\varphi_{k}+a_{-\mathbf{k}}^{\dagger}\varphi_{k}^{*}~.\end{equation}
Since the two terms of the integrand of
(\ref{falsevacuum_3point2}) are exactly the same with the
permutation of $k_2$ and $k_3$, we only consider the first term.

As can be read, (\ref{falsevacuum_3point2}) is a very complex function of three momenta.
However, since we are interested in large scales, we can expand each term of
(\ref{falsevacuum_3point2}) and only keep the lowest non-zero contribution. Then,
explicitly using the asymptotic form of the mode function $\varphi_k$ when $|\eta|\gg1$,
\begin{equation}\label{mode_asymptotic}
|\varphi_k|^2 \underset{|\eta|\gg1}{\longrightarrow} \frac{(H\eta)^2}{2\sqrt{k^2+\mu^2}}
\, ,
\end{equation}
and integrating over angles, up to linear order of $k$ we can find
\begin{align}\label{falsevacuum_3point_angle}
& \int d\Omega
|\varphi_q|^2|\varphi_{|\mathbf{k}_1-\mathbf{q}|}|^2|\varphi_{|\mathbf{k}_2+\mathbf{q}|}|^2
\left[ \mu^2 + \mathbf{q} \cdot \left( \mathbf{q} - \mathbf{k}_1 \right) \right] \left[
\mu^2 + \mathbf{q} \cdot \left( \mathbf{k}_2 + \mathbf{q} \right) \right] \left[
\mu^2+(\mathbf{q}-\mathbf{k}_1) \cdot (\mathbf{k}_2+\mathbf{q}) \right]
\nonumber\\
& = \frac{\pi}{2}(H\eta)^6  (q^2+\mu^2)^{3/2} ~ ,
\end{align}
and $k$ dependence appears only beyond quadratic order. Now we have to perform the
integration with respect to the magnitude $q = |\mathbf{q}|$. If one naively integrates
from 0 to infinity, the integral badly diverges. However, actually we should
appropriately regularize the integral as follows: we are interested in the large scale
curvature perturbation produced during the false vacuum inflation stage. Thus, we can
trace the momentum up to the scale at which false vacuum inflation ends. That is, we
should consider up to the scale $k_*$ which crosses the horizon at the moment the
potential becomes concave down. This can be found from the potential (\ref{potential}):
it becomes concave down when the effective mass is zero, i.e. $m_\phi^2 = \mu^2/a_*^2$,
which in turn gives the scale factor at this moment as $a_* = \mu/m_\phi$. The momentum
scale which crosses the horizon at this moment is thus $k_* = a_*H = (H/m_\phi)\mu$. The
modes with larger wavenumbers $k>k_*$ exit the horizon after the false vacuum inflation
phase. So for the purpose of bispectrum produced during false vacuum inflation, one
should cutoff the integration at $k=k_*$.

After imposing the cutoff at $k_*$, the integration becomes
\begin{align}
& \int d^3q
|\varphi_q|^2|\varphi_{|\mathbf{k}_1-\mathbf{q}|}|^2|\varphi_{|\mathbf{k}_2+\mathbf{q}|}|^2
\left[ \mu^2 + \mathbf{q} \cdot \left( \mathbf{q} - \mathbf{k}_1 \right) \right] \left[
\mu^2 + \mathbf{q} \cdot \left( \mathbf{k}_2 + \mathbf{q} \right) \right] \left[
\mu^2+(\mathbf{q}-\mathbf{k}_1) \cdot (\mathbf{k}_2+\mathbf{q}) \right]
\nonumber\\
& = \frac{\pi}{96} (H\eta)^6\mu^6 f\left(\frac{k_*}{\mu}\right)~,
\end{align}
where
\begin{equation}
f\left(\frac{k_*}{\mu}\right) \equiv \frac{k_*}{\mu} \sqrt{1+\frac{k_*^2}{\mu^2}} \left(
3 + 14\frac{k_*^2}{\mu^2} + 8\frac{k_*^4}{\mu^4} \right) - 3 \log \left[ 2 \left(
\frac{k_*}{\mu} + \sqrt{ 1 + \frac{k_*^2}{\mu^2} } \right) \right]~.
\end{equation}
For example, when $m_\phi=H$, we have $f(k_*/\mu) \approx 30.6318$. Thus, with the other
term of (\ref{falsevacuum_3point2}) for which we can simply replace $k_2$ by $k_3$, the
full three-point correlation function of $\calR_c$ is found to be, up to an ${\cal O}(1)$
constant,
\begin{align}\label{Phi3pt}
&\left\langle
\mathcal{R}_c(\mathbf{k}_1)\mathcal{R}_c(\mathbf{k}_2)\mathcal{R}_c(\mathbf{k}_3)
\right\rangle = \delta^{(3)}({\bf k}_1+{\bf k}_2+{\bf k}_3)
\frac{(2\pi)^7}{3A^3} \frac{H^6}{m_\phi^6 \mu^6}
f\left(\frac{k_*}{\mu}\right)~,
\end{align}
where we have used (\ref{ren_rho+p}). Comparing this with the definition of the
bispectrum
\begin{equation}
\left\langle
\mathcal{R}_c(\mathbf{k}_1)\mathcal{R}_c(\mathbf{k}_2)\mathcal{R}_c(\mathbf{k}_3)
\right\rangle \equiv (2\pi)^3\delta^{(3)}(\mathbf{k}_1+\mathbf{k}_2+\mathbf{k}_3)
B_\calR(\mathbf{k}_1,\mathbf{k}_2,\mathbf{k}_3) \, ,
\end{equation}
we can find the bispectrum of the curvature perturbation $\calR_c$ produced during false
vacuum inflation as
\begin{align}
& B_\calR(\mathbf{k}_1,\mathbf{k}_2,\mathbf{k}_3) = \frac{(2\pi)^4}{3A^3}
\frac{H^6}{m_\phi^6 \mu^6} f\left(\frac{k_*}{\mu}\right)~.
\end{align}
Note that $B_\calR(\mathbf{k}_1,\mathbf{k}_2,\mathbf{k}_3)$ is independent of ${\bf k}_i$
at linear order of $k_i/\mu$. This is very different from non-Gaussianity produced by any
other known models.

\begin{figure}
  \center
  \includegraphics[width=0.6\textwidth]{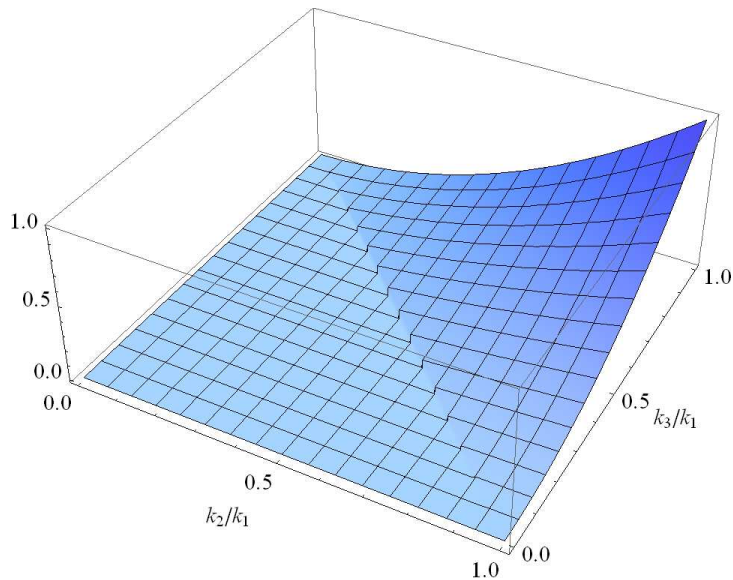}
  \caption{The shape of the bispectrum produced during false vacuum inflation. We have the
  maximum value of the bispectrum when all the momenta are of comparable magnitude.}
  \label{Fig:oldvac}
\end{figure}

In Fig.~\ref{Fig:oldvac}, we show the dimensionless shape function $(k_1k_2k_3)^2
B_\calR(\mathbf{k}_1,\mathbf{k}_2,\mathbf{k}_3)$ as a function of $k_2/k_1$ and
$k_3/k_1$. It is clear that the bispectrum exhibits its maximum value at the equilateral
limit $k_1 = k_2 = k_3$. This means, unlike the so-called local type non-Gaussianity
where non-linearity arises due to the classical super-horizon evolution of the curvature
perturbation, the curvature perturbation itself is {\em intrinsically} highly
non-Gaussian. Indeed, as we can see from (\ref{falsevacuum_Rc}), $\calR_c \propto \phi^2$
with $\phi$ being nearly Gaussian, thus $\calR_c$ follows $\chi^2$ statistics. A more
close study will follow separately~\cite{prep}.

\section{Old curvaton scenario}
\label{sec_oldcurvaton}

In the previous section, we calculated the bispectrum of $\calR_c$ produced during false
vacuum inflation. However, false vacuum inflation cannot reproduce the observed nearly
scale invariant curvature perturbation with nearly Gaussian distribution. To generate the
Gaussian contribution, we introduce a curvaton field $\sigma$. As we will see shortly, in
this ``old curvaton'' scenario we can obtain the desired nearly scale invariant power
spectrum of the Gaussian curvature perturbation, along with novel shape of the
three-point correlation function as well as large running of the spectral index.

We employ the standard curvaton scenario for producing the nearly scale invariant power
spectrum. The curvaton $\sigma$ has very small mass, no direct coupling with the inflaton
field, and sub-dominate energy density during inflation. After inflation, the curvaton
field eventually dominates the energy density either during curvaton oscillation stage or
a secondary inflation stage. Then the isocurvature perturbation stored in the curvaton
field is converted to the curvature perturbation.

In the following calculation, we will denote as ${\cal R}_\phi$ and ${\cal R}_\sigma$ the
curvature perturbations on the comoving slices of the inflaton (including possible
radiation component) and the curvaton, respectively, and use ${\cal R}_c$ to denote the
curvature perturbation on the comoving slice of the total energy-momentum tensor. As
shown in Ref.~\cite{Lyth:2004gb}, ${\cal R}_\phi$ and ${\cal R}_\sigma$ are separately
conserved quantities as long as there is no direct coupling between the inflaton $\phi$
and the curvaton $\sigma$, which is assumed in our scenario as well as in the standard
curvaton scenario.

The comoving curvature perturbation defined from the total energy-momentum tensor can be
written as
\begin{equation}\label{totalR}
  {\cal R}_c = (1-r) {\cal R}_\phi + r {\cal R}_\sigma~,
\end{equation}
where the constant $r$ is given by
\begin{equation}
  r=\left.\frac{\dot \rho_\sigma}{\dot \rho_\phi+\dot\rho_\sigma}\right|_\mathrm{dec}~,
\end{equation}
which is evaluated when curvaton decays: after then, $\calR_c$ becomes constant. From
(\ref{totalR}), the power spectrum can be written as
\begin{equation}
  \calP_\calR = (1-r)^2 \calP_{\calR_\phi}+r^2 \calP_{\calR_\sigma}~,
\end{equation}
where we have assumed no cross correlation between $\phi$ and $\sigma$. Actually, such
correlation can be induced by the gravitational coupling. But this cross correlation can
be absorbed into the redefinition of $\calR_\phi$ and $\calR_\sigma$ without modifying
any other calculation present in this paper: we will present more discussion on this
redefinition at the end of this section. The spectral index is
\begin{equation}
    n_\calR-1 \equiv \frac{d \log \calP_\calR}{d \log k}
    = 3+(n_\sigma-4)r^2\frac{\calP_{\calR_\sigma}}{\calP_\calR}~,
\end{equation}
where $n_\sigma\equiv d\log \calP_{\calR_\sigma}/d\log k$. The running of the spectral
index is written as
\begin{equation} \label{running}
  \alpha_\calR \equiv \frac{d n_\calR}{d \log k} =
  (n_\calR-1) - (n_\calR-1)^2 + 6 +
  \left[ \alpha_\sigma-(n_\sigma-1)+(n_\sigma-1)^2-6 \right]r^2
  \frac{\calP_{\calR_\sigma}}{\calP_\calR}~,
\end{equation}
where $\alpha_\sigma \equiv dn_\sigma/d\log{k}$. Note that the running of the index is of
the same order of magnitude as the spectral index. We can obtain a large running in the
old curvaton scenario.

\begin{figure}
  \center
  \includegraphics[width=0.6\textwidth]{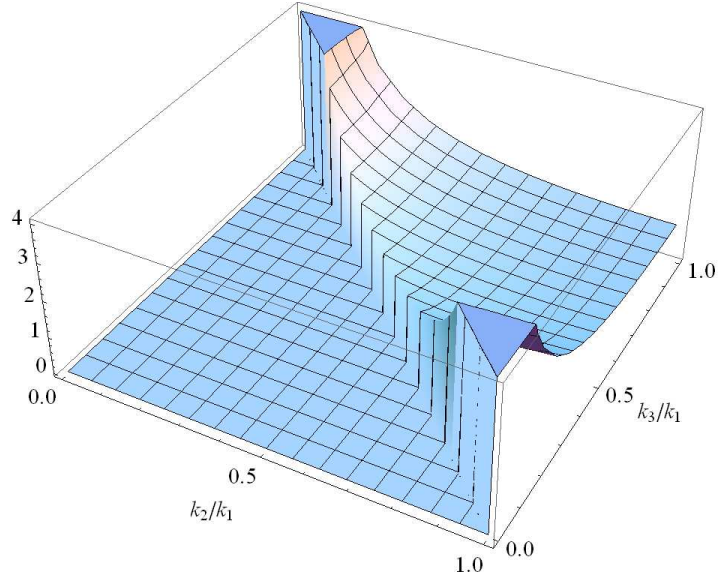}
  \caption{The local shape of bispectrum produced by the nonlinearity of curvaton. The
  bispectrum is maximum when either $k_2$ or $k_3$ is comparable to $k_1$, while the other
  one negligible.}
  \label{Fig:loc}
\end{figure}

\begin{figure}
  \center
  \includegraphics[width=0.6\textwidth]{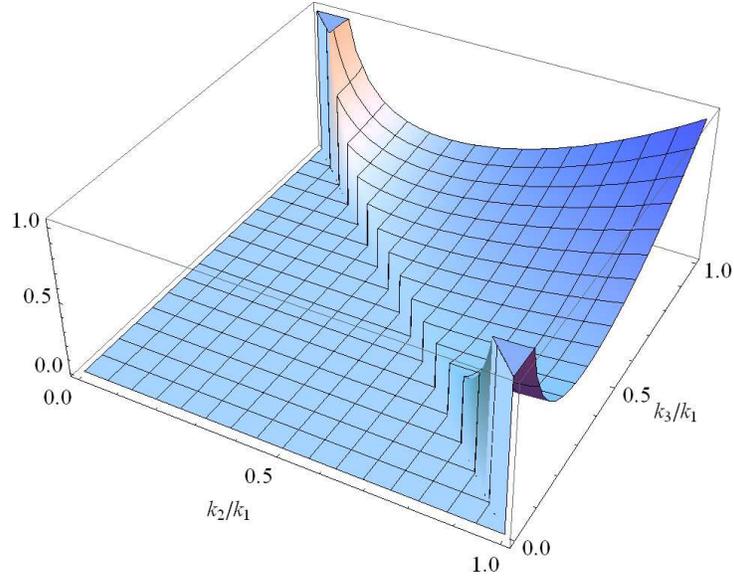}
  \caption{An example of the mixed shape between the local and the false vacuum shapes.
  The momentum dependence of the bispectrum is of the form $B_\calR(k_1, k_2, k_3) \sim
  20+\sum_ik_i/\prod_ik_i$, where the former and the latter are the false vacuum and the
  curvaton contributions, respectively.}
  \label{Fig:mix}
\end{figure}

The total three-point correlation function in the old curvaton scenario can be calculated
as
\begin{equation}\label{bispectrum}
  \langle \calR_c({\bf k}_1)\calR_c({\bf k}_2)\calR_c({\bf k}_3) \rangle
  = (1-r)^3 \langle \calR_\phi ({\bf k}_1) \calR_\phi ({\bf k}_2) \calR_\phi ({\bf k}_3) \rangle
  +  r^3 \langle \calR_\sigma ({\bf k}_1) \calR_\sigma ({\bf k}_2) \calR_\sigma ({\bf k}_3) \rangle~,
\end{equation}
with $\langle \calR_\phi ({\bf k}_1) \calR_\phi ({\bf k}_2)
\calR_\phi ({\bf k}_3) \rangle$ given by \eqref{Phi3pt}. $\langle
\calR_\sigma ({\bf k}_1) \calR_\sigma ({\bf k}_2) \calR_\sigma
({\bf k}_3) \rangle$ has a typical local shape, which is standard
in the curvaton scenario written as\cite{curvatonnG}
\begin{equation}
   r^3 \langle \calR_\sigma({\bf k}_1)\calR_\sigma({\bf k}_2)\calR_\sigma({\bf k}_3) \rangle
   = (2\pi)^7 \delta^3({\bf k}_1+{\bf k}_2+{\bf k}_3)
   \left( -\frac{3}{10} \calP_\calR^2  \right)
   \left( \frac{5}{4r}-\frac{5}{3}-\frac{5r}{6} \right) \frac{\sum_i k_i^3}{\prod_i k_i^3}~.
\end{equation}
Depending on which term is dominant in (\ref{bispectrum}), the total three-point
correlation function in the old curvaton scenario can exhibit different shape: if the
inflaton contribution is larger, the shape coincides with strong equilateral type shown
in Fig.~\ref{Fig:oldvac}, while the curvaton is dominating we have the well-known local
shape as shown in Fig.~\ref{Fig:loc}. More generally, in the old curvaton scenario in
general we have a mixed shape, lying between these two extremes. An example of mixed
shape is illustrated in Fig.~\ref{Fig:mix}.

We can also match our mixed shape non-Gaussianity with the local shape at $k_1=k_2=k_3$
to obtain a non-Gaussian estimator $\fnl^{\rm mix}$ as
\begin{equation}\label{fnl1}
  \fnl^{\rm mix} = \frac{5}{4r}-\frac{5}{3}-\frac{5r}{6} -
  (1-r)^3 \frac{10}{27A^3} \frac{k^6}{\calP_\calR^2} \frac{H^6}{m_\phi^6\mu^6}
  f\left(\frac{k_*}{\mu}\right)~.
\end{equation}
Using the expression of $\calP_{\calR_\phi}$~\cite{Gong:2008ni}, \eqref{fnl1} can be
further simplified to
\begin{align}\label{fnl}
  \fnl^{\rm mix} = \frac{5}{4r}-\frac{5}{3}-\frac{5r}{6} -
  (1-r)^3\frac{10A}{27B^2}\frac{m_\phi^2}{H^2}f\left(\frac{k_*}{\mu}\right)
  \left( \frac{\calP_{\calR_\phi}}{\calP_\calR} \right)^{2}~,
\end{align}
where $B = \mathcal{O}(0.1)$, and $m_{\phi}$ is a free parameter in our scenario. We can
get a positive and large $\fnl$ by tuning $m_{\phi}^2\gg H^2$. Note that $\fnl^{\rm mix}$
is strongly scale dependent: the scale dependence is encoded in the fraction
$\left(\calP_{\calR_\phi}/\calP_\calR\right)^2$.

%Finally, we would like to make one remark for the old curvaton
%scenario, clarifying the self-consistency of our calculation.
Finally, we would like to make some remarks on the aforementioned redefinition of the
curvature perturbations.
In the calculation, we have ignored the gravitational coupling between $\phi$ and
$\sigma$, and assume these two fields are uncorrelated. Actually, they should interact
gravitationally so that part of $\delta\sigma$ is induced from $\phi$ via $\calR_\phi$ at
second order. However, the curvature perturbation from this part of $\delta\sigma$ is
proportional to $\calR_\phi$, thus can be absorbed into the redefined $\calR_\phi$. To be
explicit, one can decompose $\calR_\sigma$ into two parts,
\begin{equation}
  \calR_\sigma = \calR_\sigma^\mathrm{(vac)}+\calR_\sigma^\mathrm{(induced)}~,
\end{equation}
where $\calR_\sigma^\mathrm{(vac)}$ originates from the vacuum fluctuation of the
curvaton, thus has nearly Gaussian distribution, while $\calR_\sigma^\mathrm{(induced)}$
is induced by the gravitation potential during the false vacuum stage of inflation, thus
follows $\chi^2$ distribution. Then the total curvature perturbation can be rewritten as
\begin{align}
\calR_c = & (1-r) \calR_\phi + r \calR_\sigma
\nonumber\\
= & (1-r) \left[ \calR_\phi + \frac{r}{1-r} \calR_\sigma^\mathrm{(induced)} \right] + r
\calR_\sigma^\mathrm{(vac)}
\nonumber\\
\equiv & (1-r) \widetilde{\calR}_\phi + r \widetilde{\calR}_\sigma \, ,
\end{align}
where we have redefined $\widetilde{\calR}_\phi$ and $\widetilde{\calR}_\sigma$. After
this redefinition, all the calculations presented in this section can be safely applied.

\section{Conclusions and discussions}
\label{sec_conclusion}

In this paper, we calculated the bispectrum of the curvature perturbation $\calR_c$
produced during false vacuum inflation. Because $\calR_c$ is intrinsically highly
non-Gaussian, the corresponding bispectrum exhibits a novel shape. Using this result, we
proposed the ``old curvaton'' scenario, where the Gaussian contribution of the total
curvature perturbation consistent with the current observations is provided by the
curvaton. The resulting shape of the bispectrum is a mixed type, i.e. the combination of
the novel ``false vacuum'' and the well known local shapes. In addition to the novel
bispectrum, in the old curvaton scenario we also find relatively large running of the
spectral index.

In the old curvaton scenario, both the inflaton and the curvaton play key roles in
producing observable effects. Moreover, there are possibly thermal components to confine
the inflaton at the top of the potential. These components set up interesting connections
between our work and some other recent progress on inflation and non-Gaussianity:

$\bullet$ {\em Quasi-single field inflation}: In the old curvaton scenario, the curvaton
field slowly rolls down its potential. One could think of this scenario in an alternative
way: to interchange the name of the curvaton and the inflaton, so that inflation is
driven by the curvaton field $\sigma$. Then our model describes a slow roll inflation
direction, plus another direction along which the field is trapped at the minimum of the
potential. In this viewpoint, our model looks similar to quasi-single field
inflation~\cite{quasiSingleFieldNG}. However, in quasi-single field inflation, the
transition from the isocurvature perturbation to the curvature perturbation comes from
direct coupling between the two directions. While in the old curvaton scenario, the
transition comes from the non-conservation of curvature perturbation after the end of
inflation, which is standard in the curvaton scenario. The origin of non-Gaussianity is
also different. In quasi-single field inflation, non-Gaussianity originates from the
interaction between the scalar fields. While in the old curvaton scenario, the
non-Gaussianity originates from the non-linear mapping from the scalar field fluctuation
to the curvature perturbation.

$\bullet$ {\em Thermal non-Gaussianity}: In the old curvaton scenario, the effective
potential \eqref{potential} can be obtained in several ways. A simple way is that the
inflaton is trapped at the origin for some $e$-folds by thermal
radiations~\cite{thermalinf}. It is possible that they also have non-Gaussianities
\cite{thermalnG}, which could be transferred to the curvature perturbation by
gravitational or other couplings. Here we note this possibility, and at the same time
emphasize that this effect is under control in our model. Thus we can turn off these
contributions, and our previous calculations give the leading non-Gaussianity.
Nevertheless, it would be interesting to see whether there are parameter regions where
these thermal effects can affect the running and the non-Gaussianity, as well as the
power spectrum and the spectral index~\cite{Gong:2006hf}.

$\bullet$ {\em Multi-stream inflation}: At the end of false vacuum inflation, the
potential \eqref{potential} becomes unstable. Then, the field could roll towards either
$\phi>0$ or $\phi<0$ at different causal patches. As discussed in multi-stream inflation
\cite{Li:2009sp}, these bifurcation does not necessarily lead to disasters, and can
rather probably give rise to some interesting observational effects, such as features,
non-Gaussianities, and the CMB asymmetries at the bifurcation scale. However, these
effects do not change the calculation and conclusion in our present paper, because they
are associated with perturbations on smaller scales.

Finally, we would like to mention that the old curvaton scenario and the related works we
discussed above all belong to the attempts assuming that the inflationary dynamics is
more complicated than the simplest single field slow roll inflation. Recent progress in
string theory, especially the string landscape~\cite{Bousso:2000xa} and related
cosmology~\cite{Huang:2008jr}, provides a theoretical playground, and also higher {\em a
priori} expectation for the complicated inflationary dynamics. Hopefully, the old
curvaton, together with a great number of other models and mechanisms, can serve as a
building block for a realistic description of inflation in the string landscape.

\subsection*{Acknowledgement}

We thank Misao Sasaki for important conversations and correspondences. JG is grateful to
the Kavli Institute for Theoretical Physics China for hospitality during the program
``Connecting Fundamental Physics with Observations'', where this work was initiated. JG
is partly supported by a VIDI and a VICI Innovative Research Incentive Grant from the
Netherlands Organization for Scientific Research (NWO). CL is supported by CSC. YW is
supported by NSERC and an IPP postdoctoral fellowship.

\appendix

\section{$\calR_c$ in terms of $\phi$ during false vacuum inflation}
\label{appendixA}

In this appendix\footnote{Some parts of this section come from Ref.~\cite{prep}.}, we
provide a little more detailed argument on how we obtained (\ref{falsevacuum_Rc}). We
know from Ref.~\cite{Gong:2008ni} that the gauge invariant energy density perturbation on
the comoving hypersurfaces $\rho\Delta$ is related to the gauge invariant intrinsic
spatial curvature perturbation $\Phi$ in the longitudinal gauge by a Poisson-like
equation,
\begin{equation}\label{Poisson}
\rho\Delta = -2\mpl^2(H\eta)^2\nabla^2\Phi \, .
\end{equation}
To extract the functional behaviour of $\Phi$, we can notice that from the energy density
correlation function,
\begin{equation}
D(x,x') = 4\mpl^4(H\eta)^4 \left\langle \nabla_x^4\Phi(x)\nabla_{x'}^4\Phi(x')
\right\rangle \sim \frac{e^{-2\mu r}}{r^3} \, ,
\end{equation}
so that $\Phi$ has an exponential factor $e^{-\mu r}$. Since we are interested in the
super-horizon separations, the Laplacian operator primarily picks the terms with the
least power of $r$ in the denominator. Thus we can approximate
\begin{equation}\label{laplacian_Phi}
\nabla^2\Phi \approx \mu^2\Phi \, .
\end{equation}
Further, $\Phi$ can be written in terms of the comoving curvature perturbation $\calR_c$
as
\begin{equation}\label{Phi_Rc}
\Phi = -\frac{A}{32\pi^2}\frac{m_\phi^2}{\mpl^2}(\mu\eta)^2\calR_c =
-\frac{\langle\rho+p\rangle_\mathrm{ren}}{2\mpl^2H^2}\calR_c \, ,
\end{equation}
where $A$ is a constant of $\mathcal{O}(1)$, and $\langle\rho+p\rangle_\mathrm{ren}$ is
the ``renormalized'' expectation value of $\rho+p$, given by
\begin{equation}\label{ren_rho+p}
\langle\rho+p\rangle_\mathrm{ren} = A\frac{H^4}{16\pi^2}\frac{m_\phi^2}{H^2}(\mu\eta)^2
\, .
\end{equation}
Therefore, combining (\ref{laplacian_Phi}) and (\ref{Phi_Rc}), from (\ref{Poisson}) we
can write $\calR_c$ in terms of the density perturbation $\rho\Delta$ as
\begin{equation}\label{Rc_rho+p}
\calR_c \approx \frac{\rho\Delta}{\langle\rho+p\rangle_\mathrm{ren}}(\mu\eta)^{-2} \, .
\end{equation}

We can move to the Fourier space more conveniently by noticing that for $D(x,x')$ the
most significant contributions come from the terms which do not contain any time
derivative. This means,
\begin{equation}
\nabla^2(\rho\Delta) = \nabla^2 \left[ \frac{\dot\phi^2}{2} + \frac{(\nabla\phi)^2}{2a^2}
+ V(\phi) \right] + 3H\dot\phi\nabla^2\phi \approx \nabla^2 \left[
\frac{(\nabla\phi)^2}{2a^2} + V(\phi) \right] \, .
\end{equation}
Further, as we are interested in the false vacuum inflation stage where
$m_\mathrm{eff}^2$ is positive, $1/a^2$ term in (\ref{potential}) completely dominates
the potential. Thus, after all, we have
\begin{equation}
\rho\Delta \approx \frac{1}{2a^2} \left[ (\nabla\phi)^2 + \mu^2\phi^2 \right] \, .
\end{equation}
Decomposing $\phi$ in terms of the Fourier mode
\begin{equation}\label{phi_Fourier}
\phi(x) = \int \frac{d^3k}{(2\pi)^3} e^{i\mathbf{k}\cdot\mathbf{x}} \phi_k(\eta) \, ,
\end{equation}
we can easily find the Fourier component of $\rho\Delta$ as
\begin{equation}\label{rhoDelta_k}
(\rho\Delta)_k = \frac{1}{2a^2} \int \frac{d^3q}{(2\pi)^3}
\phi_q\phi_{|\mathbf{k}-\mathbf{q}|} \left[ \mu^2 +
\mathbf{q}\cdot(\mathbf{q}-\mathbf{k}) \right] \, .
\end{equation}
Substituting (\ref{rhoDelta_k}) into (\ref{Rc_rho+p}), we can obtain
(\ref{falsevacuum_Rc}).

Before we finish, we note that since the potential is quadratic, we can promote $\phi$ as
a quantum harmonic oscillator and expand the Fourier mode $\phi_k$ in terms of the
annihilation and creation operators, $a_\mathbf{k}$ and $a_\mathbf{k}^\dag$ respectively,
as
\begin{equation}\label{phi_decomposition}
\phi(x) = \int \frac{d^3k}{(2\pi)^3} e^{i\mathbf{k}\cdot\mathbf{x}} \phi_k(\eta) = \int
\frac{d^3k}{(2\pi)^3} e^{i\mathbf{k}\cdot\mathbf{x}} \left[ a_\mathbf{k}\varphi_k(\eta) +
a_{-\mathbf{k}}^\dag\varphi_k^*(\eta) \right] \, ,
\end{equation}
where $a_\mathbf{k}$ and $a_\mathbf{k}^\dag$ satisfy the canonical commutation relation
\begin{equation}
\left[ a_\mathbf{k}, a_\mathbf{q}^\dag \right] =
(2\pi)^3\delta^{(3)}(\mathbf{k}-\mathbf{q}) \, .
\end{equation}
The solution of the mode function $\varphi_k(\eta)$ satisfies the asymptotic behaviour
(\ref{mode_asymptotic}) in the limit $|\eta|\gg1$.

\end{document}